\documentstyle[prl,aps,preprint]{revtex}
\begin{document}
\begin{center}
   {\Large \bf  Near-Equilibrium Dynamics  of Crystalline  Interfaces
with Long-Range  Interactions  in 1+1 Dimensional Systems:
}
\end{center}
\begin{center}
Yan-Chr Tsai
\vskip 0.25in
Department of Physics and Astronomy,
University of Pennsylvania,  Philadelphia,   PA  19104-6396

\end{center}

\begin{abstract}
The dynamics of a  one-dimensional crystalline
interface model with long-range
interactions  is   investigated.  In the absence of randomness,
the linear response mobility  decreases  to zero
when the temperature  approaches
the roughening transition  from above, in contrast
to  a finite jump
at the  critical  point in the Kosterlitz-Thouless (KT)  transition.
In the presence  of substrate disorder,  there exists
a phase transition into a low-temperature pinning phase with
a continuously  varying dynamic exponent $z>1$.
The expressions  for  the non-linear response mobility of a  crystalline
interface  in both cases are  also  derived.
\end{abstract}

\newpage

Various deposition growth models  of
interfaces  have been investigated intensively
in recent  years \cite{BN,GR1}.  The main interest
of these studies  stems from both practical
applications   and  theoretical inquires. The profile
of a  surface develops  roughness  gradually as the  stochastically
deposited   particles   accumulate.
The primary features of interface
growth are their non-trivial temporal and spatial scaling behavior.
The scaling properties are manifested in  the surface width
\cite{GR1}:
\begin{equation}
w(L,t)\sim L^{\alpha} f(t/L^{\alpha/\beta}),
\end{equation}
where $L$ is  the lateral  size of the system,
the scaling function $f(x)\sim x^{\beta} $ for
$x\ll 1$, and $f(x)\rightarrow$ constant for $x\gg 1$.
The scaling behaviors are   characterized by the roughness
exponent $\alpha$, and the  exponent $\beta$.
$\alpha$  features  the long-time limit of the self-affine
fractal structure as represented by the surface width
$w(L,t)\sim L^{\alpha}$.
At an early stage of growth, the surface width $w(L,t)$ scales as  $t^{\beta}$,
where $\beta=\frac{\alpha}{z}$
and $z$ is the dynamic exponent.

The discreteness of a  lattice is an indispensable  ingredient
in either  experiments or computer simulations.
As is  known, due to this  effect
the crystalline 2D surface   undergoes   a
roughening transition at  a finite   temperature.
Such  transitions  of various systems have been observed
experimentally  in  plastic and metal crystals.
The theoretical understanding of this
behavior is based on  the Discrete Gaussian Solid-on-Solid  model
(DGSOS) \cite{BN,CW,NG},
in which   the surface elastic  energy  is assumed to  arise   from
the interactions  between nearest
neighbor  atoms in a  crystal.
In addition,
the periodic  potential
is introduced to account for    the lattice discreteness.

One-dimensional systems  with  only short-range  interactions
should not   have a phase transition.
As a result,  the DGSOS model  in 1+1 dimension   does not exhibit a
roughening transition\cite{sosn}.
However,  in nature there exist various kinds of
long-range  interactions  between atoms,
such as the Van-der Waal force or  the effective  force due to bulk strain
\cite{Scu}.
The general form of interactions  is  found to  scale  as some  power law
with the distance  between atoms.

In view of this possibility,
we consider a particular type of surface  energy
due to  the long-rang interactions  between  atoms in
a  crystal \cite{Scu,fisher1}. This  model  could  be  realized in
vicinal surfaces whose steps are flat along one fixed direction.
Here,
we  shall  investigate the static  and dynamic properties of a  crystalline
interface  which undergoes  the  roughening
transition in one dimension.  Although  qualitatively similar to the
KT transition of the 2+1 dimensional surface, the 1+1 dimensional case
exhibits different critical behaviors.

The  2D DGSOS model with
bulk or substrate disorder has been  investigated
recently \cite{TD,TY1,sim}. The  disorder
 was found to have  non-trivial
effects on
both   the dynamical and   static
properties of the crystalline surface.
The renormalization group (RG)  analysis predicts
 a
so-called super-roughening transition.
The low temperature phase is dominated  by  disorder, and displays
 glassy behavior, identical to the vortex glass phase
found  in   2-dimensional dirty type-II
superconductors \cite{FFH89,nat,hwa1,rsb}.
Here we shall incorporate  the same type of disorder into the 1D interface
with the elastic energy due to long-range  interactions.

The motivation  of   this study  is
twofold.  First,  we would like to
investigate   the behavior  of an  interface
with the  elastic energy  arising from long-range  interactions  in
two different environments- -pure and disordered.
Secondly,
current computer simulations  of
an  2D interface   subject to  quenched disorder
are limited to modest   spatial and temporal scales,
and the present model could  be used to   probe more effectively  the
hydrodynamic limit.

The  elastic  energy of the interface caused by
the long-range  interactions  \cite{Scu,fisher1} considered here
is:
\begin{equation}
{\cal H}_0 = \frac{\rho}{2 T}\int_{-\infty}^{\infty}  dx
 \int_{-\infty}^{\infty
}   dx^{'} \Bigl[ \frac{h(x)-h(x^{'})}{x-x^{'}}\Bigr]^2
=\frac{K}{2}\int_{-\infty}^{\infty}\frac{dp}{2\pi} |p| h(p)h(-p),
\label{eq:long}
\end{equation} where $h(x)$ is the height of the interface at
position $x$, and  $\rho$ is the
strength characterizing the elastic energy  due to
long-range  interactions between atoms.  The temperature
$T$ has been absorbed into
the Hamiltonian,
and $K=2\pi \rho /T$. Powercounting shows that surface energy due to
short-range  interactions
$\int dx (\frac{dh}{dx})^{2}$  is irrelevant. Therefore
we will not take  it into account here.

The periodic potential inherited
from  the discrete nature of lattice is  given by:
\begin{equation}
{\cal H}_{\text{p}}= \frac{\alpha}{a} \int _{-\infty}^{\infty}   dx  \cos
[\gamma h(x)],
\label{eq:per}
\end{equation}
where $\frac{\alpha}{a}$
is the  coefficient of
the leading term due to
the discreteness, and
 $\gamma= \frac{2\pi}{b}$ with $a$ and $b$ being   lattice
constants along  and perpendicular to the base line, respectively.
With this potential, the height of a  surface should be regarded as
a continuous variable within the interval $(-\infty,\infty)$.
As in  the DGSOS model,
the higher harmonics are  irrelevant near the critical point  and
can therefore  be  neglected.

The action in Eq.~(\ref{eq:long})
can  also  describe a variety of
systems with quantum dissipations.
A particular case
is the   electronic  system involving  a quantum wire with one impurity
scattering center
\cite{Kane1}.
The Euclidean  action of the electron at
the impurity center is equivalent  to the sum of    Eq.~(\ref{eq:long})
and Eq.~(\ref{eq:per}).

The dynamics of interface  growth considered
here is governed by\cite{CW,NG,Neu}:
\begin{equation}
\frac{\partial h(x,t)}{\partial t} =\mu  F - \mu  \frac{\delta {\cal H}}
{\delta h}
+\eta(x,t),
\label{eq:mo}
\end{equation}
where $t$ is time, $F$ is the driving force,
$\mu$ is  the mobility,  and  ${\cal H}$
is  the sum of  ${\cal H}_{\text{0}}$ and ${\cal H}_{\text{p}}$.
$\eta$  is the thermal  noise which obeys
$\langle\eta(x_1,t_1)\eta(x_2,t_2)\rangle=2 \mu
 \delta(x_1 - x_2) \delta (t_1-t_2)$,
where  the coefficient  $2\mu$
is chosen  such that the system will
reach the equilibrium states  weighted  by the
Boltzmann  factor  $e^{-(H_0+H_p)}$.
Consequently  the fluctuation-dissipation  theorem \cite{Zinn} will hold here
in
contrast to the case of far-from equilibrium surface growth \cite{KPZ}, which
cannot be formalized  by Eq.~(\ref{eq:mo}).

To facilitate the analysis of Eq.~(\ref{eq:mo})
 we  employ the  formalism of
Martin, Siggia, and Rose (MSR) \cite{MSR,janssen}. In addition to
$h(x,t)$, the auxiliary field $\tilde {h} (x, t) $ is
introduced as well as their conjugate source $J(x,t)$ and $
\tilde{J} (x,t)$. Upon averaging over the thermal noise
the associated generating functional is obtained:
\begin{equation}
Z[J,\tilde{J}]=\int {\cal D}h {\cal D}\tilde{h} e^{{\cal S} _{\text{eff}}}
=\int {\cal D}h {\cal D}\tilde{h} e^{{\cal S} _{\text{0}}+{\cal
S}_{\text{int}}},
\label{eq:gen}
\end{equation}
where the effective MSR action  ${\cal S}_{\text{eff}}$ takes the form:
\begin{eqnarray}
{\cal S}_{\text{0}}&=& \int _{-\infty}^{\infty} \frac{dp}{2\pi}
\int _{-\infty}^{\infty}
dt \{J h +\tilde{J} \tilde{h}+
\mu \tilde {h}(p,t) \tilde h(-p,t)
- \tilde h(-p,t) [\frac
{\partial}{\partial t}  h(p,t) -K\mu |p| h(p)]\}  \label{eq:s0}, \\
{\cal S}_{\text{int}}&=& - \frac{\mu \gamma \alpha}{a} \int _{-\infty}^{\infty}
dx \int _{-\infty}^{\infty}
dt  \tilde h(x,t)  \sin[\gamma h(x,t)].
\label{eq:si}
\end{eqnarray}
Here   Eq.~(\ref{eq:gen})	will be treated   by the RG technique.
To perform the RG calculation, we follow the schemes of
Amit  ${\it et}$  $\it{al}$.\cite{SGT} and  Neudecker \cite{Neu}.
Regularization of   the Feynman integrals is
established    by introducing
the infrared cutoff $M$ and ultraviolet cutoff $a$ into the
free correlation function $C_{0}(x,t)
= \langle h(x,t)h(0,0)\rangle =\int _{-\infty}^{\infty} \frac{dp}{2\pi}
\frac{1}{K|p|+M} e^{-\mu (K|p|+M)|t| } e^{-ip y}
    $with $y^2=x^2+a^2$.

In  the absence of  the periodic
potential,  the 1+1 dimensional interface width  scales  as \cite{nat1}:
\begin{equation}
w^2(L,t)  \sim \frac{\mu}{K}\ln\Bigl[ \frac{L}{a}f\Bigl( K\frac{t}{L}\Bigr)
\Bigr],
\label{eq:frcf}
\end{equation}
where $L$ is the linear size of the system, and the scaling
function $\ln f(x)=-\ln2 +\text{Ei}(4\pi x)$ with $\text{Ei}(-y)=-\int _{y}
^{\infty}du \frac{e^{-u}}{u}$.

\underline {I. Roughening transition in 1+1 dimensional systems}:

To analyze Eq.~(\ref{eq:gen})  we  calculate
the perturbative correction to several  vertex functions
up to the lowest order in $\tau=1-\frac{\gamma ^2}{2\pi K} $
and  the second order in  $\alpha$, and use the minimal subtraction
and the standard prescription to obtain the  flow equations of various
parameters \cite{Zinn,amit} (details of the calculation will be published
elsewhere).

Under rescaling factors $x\rightarrow e^{l} x $ and
$t\rightarrow t e^{l z}$ ($l=\ln  y  $ where $y$  is the rescaling
factor), the  RG recursion relations of various  parameters are given by:
\begin{eqnarray}
\frac{dK}{dl}&=&0 , \label{eq:1rk}  \\
\frac{dF}{dl}&=&F , \label{eq:1rf}  \\
\frac{d\alpha}{dl}&=&(1-\frac{\gamma ^2  }{2\pi K}) \alpha +O(\alpha ^3),
\label{eq:1ra}  \\
\frac{d\mu}{dl}&=&\Bigl(z-1-  ( \pi \alpha)^2\Bigr)\mu \label{eq:1rm} .
\end{eqnarray}
We first  discuss static properties.
Eq.~(\ref{eq:1rk})  holds true for all orders  of $\alpha$, since there are  no
conterterms proportional to $|p|$ in any order.
Note that under the RG transformation
only the terms proportional  to  powers of $p$, i.e. local in space,
can  be generated.
Non-renormalization
of $K$ is also justified by using the  real space RG calculation \cite{Kane1}.
On the other hand,  in the case of   2D crystalline  interface
the surface tension  does suffer  renormalization.
As shown in Eq.~(\ref{eq:1ra})  the sign of $\tau=
1-\frac{\gamma ^2}{2\pi K}=1-\frac{T}{T_c}$,
determines the fate of  coupling constant $\alpha$, where $T_c=b^2\rho$.
When $\tau <0$ or $T>T_c$,
the flow  of $\alpha $ under the RG transformation was driven to
zero, and thus   lattice effect  can be neglected.
In  the long-wavelength limit,
the surface  width $w(L, t=\infty)$  scales  as $(\ln L)
^{1/2}$, as implied from Eq.~(\ref{eq:frcf}).
On the other side of the transition($T<T_c$), as argued in  Ref. 7,
under the RG iteration  coupling constant $\alpha$  flows to the infinity.
On  a sufficiently  large length scale, a  mass term  generated
by the cosine term  will dominate   over the surface energy strength $K$.
The interface is  pinned at  the  minima of the periodic
potential, and is led to a flat phase as  also occurs in 2+1
dimensional systems. Although there is no fixed point of $\alpha$
for $\tau >0$, $\alpha(l)=\alpha(0) e^{\tau l}$ should hold
as long as $l\ll 1/\tau = l_D $.

Next we turn   to the  dynamical properties of the interface in both
phases. To facilitate the discussion, we give the
expression of $\mu$ on  the  length scale $l$   by integrating
Eq.~(\ref{eq:1ra}) and
Eq.~(\ref{eq:1rm}):
\begin{equation}
\mu(l)=\mu(0) e ^{-\frac{\pi ^2 \alpha(o) ^2}{2\tau}
(e^{2\tau l}-1)}.
\label{eq:1mur}
\end{equation}

The
linear response mobility (as  $F\rightarrow 0$)
can be  obtained  by letting $l=\infty$ in Eq.~(\ref{eq:1mur}).  We find:
\begin{equation}
\mu _M = \mu(0) e^{\frac{ \pi ^2\alpha(0) ^2}{2 \tau}}
\end{equation}
for $\tau <0$ and $ \mu_M =0 $ for $\tau >0$. Here
the  parameters followed by  $(0)$
represent   the corresponding bare values.
In contrast to the roughening transition in the 2+1 dimension case,
the linear response mobility  decreases   to zero as the  temperature
approaches the   transition point from above.
The higher order correction will not   change
this behavior  qualitatively.
In the case that the applied force $F$ is finite, the movement
of an interface is characterized  by the  non-linear response mobility.
Since $F$  is a relevant field, on  some sufficiently  large
length    scale $l^{*}\sim -\ln F$, it will grow
significantly. The pinning effect will be  wiped out
by the motion of an  interface.
Beyond this scale,
the effect of the strength $\alpha$, on the average,
can be neglected.
The mobility $\mu$
will scale normally, and  its
canonical dimension  is zero.
Consequently,  the non-linear response mobility is  yielded by stopping the
RG iteration up to this scale:
\begin{equation}
\mu(l^{*}\sim -\ln F) \sim \mu(0) \exp \Bigl[- \frac{C}{\tau F^{2 \tau}}\Bigr],
\quad \quad \tau > 0
\label{eq:nonm}
\end{equation}
where $C$ is a  constant depending on bare values of
other parameters. Meanwhile $l^{*}$ shoould be less than $l_D$  i.e. $-\ln F  <
\frac{1}{\tau}$.
 Such a form of nonlinear response has been of interest to statics of the
vortex glass phase of type II superconductors. Here we see it can arise even
in pure systems.

In the region far below the transition point
the results presented here in general will not hold. However,
one can resort to the activated-dynamics mechanism.

\underline{II. Depinning transition  in 1+1 dimension}:

Crystalline surface growth
with  bulk disorder or  two-dimensional   substrate disorder
has been   addressed  recently.
Here we shall  study  the dynamics of a 1+1 dimensional
interface  with  elastic energy in  Eq.~(\ref{eq:long})
subjected to  substrate  disorder.

In the presence of substrate
disorder,   the periodic potential
\cite{TY1}
changes to:
\begin{equation}
{\cal H}_{\text{dis}}= \frac{\alpha }{a}\int dx  \cos (\gamma [h(x)+ d(x)]),
\end{equation}
where $d(x)$  is  the quenched fluctuation of the substrate height
uniformly
distributed within  the lattice spacing $b$.
The correlation of $d(x)$  under question is presumed to be
of short-range  within
lattice spacing $a$. In terms of the phase-like variable $\theta(x)=
\gamma d(x)$, we assume:
\begin{eqnarray}
\langle e^{i\theta(x)}\rangle &=&0 ,\label{eq:dis1}  \\
\langle e^{i\theta(x)}e^{-i\theta(x^{'})}\rangle &=&a \delta(x-x^{'}).
\label{eq:dis2}
\end{eqnarray}

As can be seen in Eq.(17), substrate disorder acts  to destroy the structure of
a  regular
lattice, thereby eliminating the roughening transition.
However, as we show below,
the  correlation  of  disorder will lead
to a new term which is relevant  at  lower  temperatures.

After averaging over disorder $\theta$,
the corresponding MSR action  becomes:
\begin{equation}
S_{\text{int}}=
\frac{\mu ^2 g \gamma ^2}{2a} \int _{-\infty}^{\infty}dx  \int
_{-\infty}^{\infty}  dt
\int _{-\infty}^{\infty}  dt^{'} \tilde{h}(x,t) \tilde{h} (x,t^{'})
\cos[\gamma h(x,t)-\gamma  h(x,t^{'})],
\label{eq:dis}
\end{equation}
where $g(0)=\frac{\alpha ^2(0)}{2}$, and  again higher harmonics are discarded.

Now we apply the RG analysis to the MSR action combining Eq.~(\ref{eq:s0})
and  Eq.~(\ref{eq:dis})
in  the same spirit as  we   did
the previous case.
For the present theory the expansion parameters are
$\tau ^{'}=1-\frac{\gamma ^2}{K \pi}$ and $g$.
We derive the  recursion relations for $x\rightarrow xe^{l}$, and
$t\rightarrow te^{lz}$ as following:
\begin{eqnarray}
\frac{dK}{dl}&=& 0 , \label{eq:2rk} \\
\frac{dF}{dl}&=&F , \label{eq:2rf} \\
\frac{dg}{dl}&=& (1-\frac{\gamma ^2}{K \pi} )g - 2g^2 , \label{eq:2rg}\\
\frac{d\mu}{dl}&=&(z-1 - \pi  g) \mu  \label{eq:2rm}.
\end{eqnarray}

As inferred from Eq.~(\ref{eq:2rg}) above,  the transition point is
$T_{pc}=b^2\rho /2$,
i.e.  $\tau ^{'} =1-\frac{T}{T_{pc}}<0$.
The coupling constant $ g$ pertaining  to
disorder flows  to zero under the RG transformation.
Below $T_{pc}$,  $g$ flows to a non-Gaussian fixed point
$g^{*}=\frac{\tau^{'}}{2}$.

As with the previous case,
we first discuss the static properties.
In the high temperature phase,  disorder is irrelevant and
 the surface width (or static height-height  correlation function)  is
$w(L,t=\infty)\sim (\ln L)^{1/2} $.
In contrast to the random DGSOS model, the marginal
time-persistent parts \cite{Domi,GS,TY2}
such as  $\int \int\int  dp dt dt^{'}|p| \tilde{h}(p,t)\tilde{h}(-p,t^{'})$
are  not generated,  since the term proportional to $|p|$ is spatially
nonlocal.
The form of the static correlation function
will rely on the existence of such a term, as
 in  the 2D case. Near the transition point
the height-height correlation functions
of two phases   will not be  of any difference in contrast to the  2D random
DGSOS
model.

Next  we turn to the  dynamical properties. Above $T_{pc}$,
$z=1$, and the linear response mobility remains finite and decreases
  to   zero as  the temperature approaches  the transition point. The
integration of the RG recursion relations gives the expression of
the  mobility at the scale $l$:
\begin{equation}
\mu(l)=\mu(0) [1+\frac{2g(0)}{\tau ^{'}}(e^{\tau ^{'}l} -1)]^{-\pi/2}.
\end{equation}
and yielding  the linear response mobility:
\begin{equation}
\mu _{M} = \mu_0 \frac{|\tau ^{'}|^{\pi/2}}{[2 g(0)]^{\pi/2}} ,
\quad \quad \text{as}\quad  \tau^{'}  \rightarrow 0^{+}.
\end{equation}

Below $T_{pc}$, due to the existence of  an infrared stable fixed point
$g^{*}$, the dynamic exponent $z$ assumes a non-trivial value
$1+\frac{\pi}{2}(1-\frac{T}{T_{pc}})$. The linear response mobility
vanishes below $T_{pc}$.

Along  the same line  we derived   Eq.~(\ref{eq:nonm})
the  nonlinear response mobility below $T_{pc}$ is  given by:
\begin{equation}
\mu  \sim \mu(0) F^{\frac{\pi}{2}|\tau^{'}|}
\end{equation}
to leading order in $\tau^{'}$.

The physical implications of  the RG analysis in the
low-temperature phase are  the following:
Unlike  the case of the  2D  interface,
the time persistent term in the 1D  case  is not
generated under the RG transformation. Physically
the interface with long-range  interactions   cannot afford
costly  elastic energy
to adjust itself to  follow
 the short-range  correlated substrate disorder.
In the low temperature phase   of the 2+1 random DGSOS model,
the surface does follow   disorder.
In the present case, disorder is also
relevant in the low temperature phase.
In the presence of  a force,
the  traveling interface on the average is attracive
to  the local minima of
the random periodic potential.
This implies a larger dynamic exponent
and a  smaller  mobility  in low-temperature phase than in the other
phase.

It is straightforward to extend the calculations for the long-range
elastic energy with  other powers in the denominator of  Eq.~(\ref{eq:long}).
 The  found behaviors  are  generic to  systems  with long-range
interactions.  The  qualitative  difference  between   systems
 with  short-  and long-range
interactions   is attributed  to whether or not
 the  interactions are renormalized.

In conclusion, we have investigated the dynamics of a 1+1 dimensional
crystalline interface with long-range  interactions. The governing equation
is cast in terms of the MSR formalism and is   analyzed
by using the RG technique.
In the presence  of a  regular lattice, above $T_c$,   the interface width
is $w(L,t)\sim \Bigl [\ln \Bigl(Lf(\frac{t}{L})\Bigr )\Bigr ]^{1/2}$, and  the
linear-response mobility  is
$\mu _M = \mu(0) e^{\frac{ \pi ^2\alpha(0) ^2}{2 \tau}}$.
Below $T_c$ the interface is
flat.  The  nonlinear mobility  is
$\mu \sim \mu(0) \exp \Bigl[- \frac{C}{\tau F^{2 \tau}}\Bigr] $.

In the presence of  substrate disorder,  above $T_{pc}$
 the interface width is
$w(L,t)\sim \Bigl [\ln \Bigl(Lf(\frac{t}{L})\Bigr )\Bigr]^{1/2}$.
Below $T_{pc}$,  the surface width is
$w(L,t)\sim \Bigl[\ln \Bigl(Lf(\frac{t}{L^z})\Bigr)\Bigr]^{1/2}$ with
$z=1+\frac{\pi}{2}(1-\frac{T}{T_{pc}})$ .
The expression  of the nonlinear mobility is also   obtained   for
disordered   case.

{\bf Acknowledgments}

The author  is
indebted to  T. Hwa  for  critically reading the manuscript and bringing
insightful suggestions.
He also would like to thank  to  C. Kane, T. Lubensky, Y. Shapir and H. Stark
for useful discussions.

\end{document}